# Combination of Raman spectroscopy and chemometrics: A review of recent studies published in the *Spectrochimica Acta, Part A: Molecular and Biomolecular Spectroscopy* Journal

Yulia A. Khristoforova*, Lyudmila A. Bratchenko, and Ivan A. Bratchenko

Samara National Research University, 34 Moskovskoye shosse, Samara 443086, Russia

* corresponding author khristoforovajulia@gmail.com

**Abstract**

Raman spectroscopy is a promising technique used for noninvasive analysis of samples in various fields of application due to its ability for "fingerprint" probing of samples at the molecular level. Chemometrics methods are widely used nowadays for better understanding of the recorded spectral "fingerprints" of samples and differences in their chemical composition. This review considers a number of manuscripts published in the *Spectrochimica Acta, Part A: Molecular and Biomolecular Spectroscopy* Journal that presented findings regarding the application of Raman spectroscopy in combination with chemometrics to study samples and their changes caused by different factors. In 57 reviewed manuscripts, we analyzed application of chemometrics algorithms, statistical modeling parameters, utilization of cross validation, sample sizes, as well as the performance of the proposed classification and regression model. We summarized the best strategies for creating classification models and highlighted some common drawbacks when it comes to the application of chemometrics techniques. According to our estimations, about 70% of the papers are likely to contain unsupported or invalid data due to insufficient description of the utilized methods or drawbacks of the proposed classification models. These drawbacks include: (1) insufficient experimental sample size for classification/regression to achieve significant and reliable results, (2) lack of cross validation (or a test set) for verification of the classifier/regression performance, (3) incorrect division of the spectral data into the training and the test/validation sets; (4) improper selection of the PC number to reduce the analyzed spectral data dimension.

## 1. Introduction

Raman spectroscopy (RS) is an advantageous spectral method that provides "fingerprint" information on the biochemical composition of a tested sample [1]. RS is characterized by high chemical specificity and covers a wide range of application tasks regarding the sample chemical composition in various fields such as biological, medical, pharmaceutical, and environmental applications. RS can be applied for studying the influence of external factors on the samples [2, 3], assessing the composition quality of the samples [4], identifying the pathogenic bacteria [5], developing disease diagnostic tasks [6], theranostics tasks [7], etc.

It is well-known, that a Raman spectrum of one sample consists of the contribution of the Raman signals of all active molecules. Changes in the sample chemical features caused by different factors do not include the appearance of additional Raman peaks only , but also the changes in the intensities and widths of the Raman bands that are typical for the sample under study. For spectral features, qualitative analysis cannot be performed, therefore, high RS sensitivity to chemical features at the molecular level requires special mathematical tools to reveal and retrieve important information.

The use of chemometrics methods has become widespread for Raman data analysis, which is clearly reflected in a large number of manuscripts in the last few years [8,9]. These papers deal with important differences in the specimens' biochemistry, because the combination of RS and the chemometrics methods is able to build statistical models for recognizing concentrations or group information on specimens using their spectral data. Numerous research papers contain analysis of Raman data using Principal Component Analysis (PCA) [10], Projection on the Latent Structures (PLS) [11, 12], Hierarchical Cluster Analysis (HCA) [13], Support Vector Machine (SVM) [14], and Convolution Neural Network (CNN) [15], etc. However, in-depth study of these papers has not reveal any generalized aspects when the chemometrics methods are applied to the analysis of spectral data and construction of statistical models.

This raises the question of proper use of chemometrics to obtain reliable results. We have reviewed the papers that deal with the application of chemometrics methods for analyzing Raman spectral datasets. Further, we summarized the successful approaches and highlighted specific issues in the analyzed studies. This information helps find the most appropriate and correct approaches to obtain reliable results in Raman spectral data classification and, at the same time, to point out typical errors regarding the application of chemometrics methods for building classification models.

## 2. Methods

For this review, we have analyzed the papers published in the *Spectrochimica Acta, Part A: Molecular and Biomolecular Spectroscopy* Journal from Volume 244 (2021) to Volume 267 (2022). We primarily focused on the papers dealing with the application of RS technologies including conventional RS and SERS techniques for different classification tasks using machine learning and chemometrics methods. We analyzed papers with different samples: different forms of biological samples (biotissues, biofluids, bacteria and pathogens forms, plant species), pharmaceutical drugs, and chemical samples. In total, our critical review includes 57 papers that are presented in Table 1.

In these papers, we evaluated the stability of the proposed chemometrics models (including such parameter as the number of utilized principal components or loadings), cross validation application of cross validation, sample sizes, the classifier and regression performance. By analyzing these aspects, we were able to determine the correctness of the proposed models and the possibility to further develop a reliable classifier based on these models.

## 3. Results and discussion

Table 1 summarizes the analyzed papers and contains our short comments on the chemometrics techniques utilized for the analysis of Raman spectra based on the details presented by the authors in their papers. The reviewed papers cover a wide range of applications utilizing different chemometrics approaches. Thus, the PLS/PLS-DA was performed in 30 studies (53%), the PCA/PCA-LDA – in 21 studies (37%), and the neural networks were utilized in 6 studies (10%). On the ground of the revealed potential pitfalls of the proposed models (that are summarized in Table1) and lack of complete description of the utilized chemometrics techniques, only about 30% of the presented studies (17 out of 57) can be considered capable of demonstrating a fully reliable classification and regression model without possible drawbacks.

Table 1 General information of reviewed papers

| № | Author | Techniques | Comments | Conclusion |
|---|---|---|---|---|
| 1 | X. Chen et al. [15] | PCA+kNN, PCA+LDA, PCA+SVM | The LOO CV is used; 30 spectra are recorded and analyzed for each serum sample; the spectra of the same sample might have been included in both the training and the test sets. The shape of the PCs is not shown. | The model is probably incorrect. |
| 2 | A. H. Arslan et al. [13] | PCA + HCA | The data regarding the CV results are missing. The shapes of only two first PCs are shown, whereas the first ten PCs were fed into the HCA. | There are no detailed data on the proposed model. |
| 3 | M. M. da Mata et al. [16] | PLS-DA | Several spectra were recorded and analyzed for each sample, the spectra of the same sample might have been included in both the training and the test sets. | The model is probably incorrect. |
| 4 | I. Behl et al. [17] | PLS-DA | The CV method is adequate. Several spectra were recorded and analyzed for each sample. However, the LOPO CV is used, therefore the model was trained and tested on different data. | The model is correct. |
| 5 | S. Gao et al. [12] | PLS-SVM | The experimental data on the training and the test sets were correctly divided; the 10-Fold CV was used. | The model is correct. |
| 6 | Y. Zhu et al. [10] | PCA – SWLDA | The sample size is too small to draw practical conclusions about the model. | The results are probably unreliable. |
| 7 | M. Ye et al. [18] | PC-CRT | There is no mention of the method of selecting the PC number to train the model. | There are no detailed data for the proposed model. |
| 8 | N. Iturrioz-Rodríguez et al. [19] | PCA-LDA | Only 1 sample per specific group. | The model is probably incorrect. |

| | | | 15 spectra were recorded and analyzed for each sample and cell type; the spectra of the same sample might have been included in both the training and the test sets. | |
|---|---|---|---|---|
| 9 | C. Chen *et al*. [20] | Alex Net, MCNN | The spectra of the same sample might have been included in both the training and the test sets. | The model is probably incorrect. |
| 10 | M. N. Ashraf *et al*. [21] | PCA | There is no detailed description of the methods used for data analysis. It is difficult to assess the correctness of the proposed model. | There are no detailed data on the proposed model. |
| 11 | Q. Liu *et al*. [22] | GA-SVM, RF, BPNN | Independent validation set was used as an important reference for evaluating the effectiveness of models | The model is correct. |
| 12 | A. Zhu *et al*. [23] | PLS-DA | There is no detailed description of the methods used for data analysis. It is difficult to assess the correctness of the model. | There are no detailed data on the proposed model. |
| 13 | T. Dou *et al*. [24] | PLS-DA | There is no detailed description of the methods used for data analysis. It is difficult to assess the correctness of the model | There are no detailed data on the proposed model. |
| 14 | S. Shafaq *et al*. [4] | PLSR | LOSOCV was used. | The model is correct. |
| 15 | B. Li *et al*. [25] | PCA-LDA | Despite the use of LOPOCV, 15 PCs were utilized to build model for a dataset of 220 spectra, therefore the model is overfitted; a large decrease in accuracy of the test set (75%) compared to the training set (90%) may be due to model overfitting. | The model is probably overfitted. |
| 16 | J. Depciuch *et al*. [26] | PLS, PCA, HCA | There is no detailed description of the number of samples and methods utilized for data analysis. Thus, it is impossible to assess the correctness of the model. | No detailed data for the proposed model. |
| 17 | J. C. Ramirez-Perez *et al*. [27] | PCA | The PCA is used only to demonstrate the differences between the spectral features of different groups. | There are no statistical model is provided |
| 18 | Y. Lin *et al*. [28] | PCA-LDA | The 10-Fold CV was used. The authors used the mean squared error of prediction (MSEP), as a more statistically value method for selecting the number. | The model is correct. |
| 19 | M. Kopec *et al*. [29] | PLS-LDA | It is difficult to understand the statistical modeling, because the authors refer to the previous work that studied different experimental samples. For each sample type, thousands of spectra were recorded and analyzed, therefore the spectra of the same sample might have been included in both the training and the test set. | There are no detailed data for proposed model. |
| 20 | X. Zhao *et al*. [2] | PLS | Since CV was not used, the accuracy might have significantly decreased in the test set for any number of PCs. | The model is probably incorrect. |
| 21 | M. Kashif *et al*. [5] | PCA, PLS-DA | There is no description of the CV, the training or the test sets, therefore the model might be incorrect. It is not clear why 16 LVs were utilized to build the model; the first two LVs can differentiate the analyzed samples according to the presented score plot | There are no detailed data for the proposed model. |
| 22 | J. Lei *et al*. [30] | PCA, PLS-DA | There is no description of the statistical analysis in the Methods section, therefore, it is not clear why the authors utilized three PCs. | There are no detailed data for the proposed model. |
| 23 | Y. Lin *et al*. [31] | PLS-DA | The LOO CV method is used but the results for CV utilized in the model are not provided. | There are no detailed data for the proposed model. |
| 24 | L. Jiang *et al*. [32] | GA-PLS, UVE-PLS, | No CV is used. | The model is probably |

|    |                         | VCPA-PLS<br>CARS-PLS |                                                                                                                                                                                                                                                                                                                                                                               | overfitted.                                                       |
|----|-------------------------|----------------------|-------------------------------------------------------------------------------------------------------------------------------------------------------------------------------------------------------------------------------------------------------------------------------------------------------------------------------------------------------------------------------|-------------------------------------------------------------------|
| 25 | S. Zhu et al. [33]      | PLS-DA               | Ten loadings are used to train the model, but the loadings from 3 to 10 have significantly smaller amplitudes in comparison with loading 1 and loading 2 and can be associated with noise components.<br>The model is likely overfitted.                                                                                                                                       | The model is probably overfitted.                                 |
| 26 | Q. Wang et al. [34]     | NIPALS               | The authors did not indicate the number of PCs used to build a PLS model.                                                                                                                                                                                                                                                                                                     | There are no detailed data for the proposed model.                |
| 27 | G. Orilisi et al. [35]  | PCA                  | The sample size is too small to draw practical conclusions about the model.                                                                                                                                                                                                                                                                                                   | The result is probably unreliable.                                |
| 28 | S. Rafiq et al. [36]    | PCA, PLS-DA          | 15 LVs are used to build the model, but, according to the presented 2D scores plot, the first LV can discriminate the analyzed samples with high accuracy (about 100%). It is not clear why the authors utilized 15 LVs in the model.                                                                                                                                         | There are no detailed data for the proposed model.                |
| 29 | H. Wang et al. [37]     | PCA-LDA              | Seven PCs are used to build the model, but, according to the presented 3D scores plot, the first three PCs can discriminate the analyzed samples with high accuracy (about 100%). It is not clear why the authors utilized seven LVs in the model.<br>The authors used the CV method but did not provide the RMSE or RMSECV data. | There are no detailed data for the proposed model.                |
| 30 | M. Bonsignore et al. [38] | PCA                | All the data about model are provided.                                                                                                                                                                                                                                                                                                                                        | The model is correct.                                             |
| 31 | S. Bashir et al. [39]   | PLS-DA               | The authors utilized 14 LVs to train the PLS model, but, according to the scores plot, the samples are clearly differentiated using two LVs. Using more than two LVs                                                                                                                                                                                                          | The model is probably overfitted.                                 |

| | | | leads to model overfitting. | |
|---|---|---|---|---|
| 32 | R. Dong et al. [40] | DASN | 6000 spectra for 300 samples were recorded and divided into the training and the test sets. Therefore, the spectra of the same sample might have been included in both the training and the test sets that leads to invalid model estimation. | The model is probably overfitted. |
| 33 | H. Guan et al. [41] | PCA, PLS-DA | The authors selected seven PCs to train the PLS model using the minimal RMSEP, but this condition is not sufficient for selecting the correct number of PCs. RMSECV should only complement the analysis of correct PC number. | The model is probably overfitted. |
| 34 | D. Ma et al. [42] | CNN | 15 spectra were recorded for each of the 20 samples and randomly divided into the training and the test sets. The spectra of the same sample might have been included into both the training and the test sets that leads to invalid model estimation. | The model is probably incorrect. |
| 35 | M. Roman et al. [3] | PLSR | The authors selected the number of LVs to train the PLS model on the basis of the minimal CV error. However, the model may be overfitted as it should be compared with the discrepancy of the model error without the CV error. In addition, the shape of the loading should have been estimated. | The model is probably overfitted. |
| 36 | F. Batool et al. [43] | PLS-DA | The authors selected 13 LVs to train the PLS model on the basis of the minimal RMSECV, but for 11 LVs a discrepancy between the error for the training and the CV data is observed. So, using more than 11 LVs to | The model is probably overfitted. |

| | | | train a PLS model is not correct. Moreover, the shape of the loadings should be estimated, because loadings of high order may be associated with noise components. | |
|---|---|---|---|---|
| 37 | N. M. Ralbovsky et al. [44] | PLS-DA | The presented results are reliable, because the true performance of the model was estimated using external validation of the new experimental data unseen by the model. | The model is correct. |
| 38 | C. He et al. [45] | SVM | 30 spectra were recorded for each of the 54 patients. The authors did not describe the method of splitting the spectra into the training and the test sets. The spectra of the same patient might have been included into the training and the test sets, which can lead to the invalid model. | The model is probably overfitted. |
| 39 | C. Robert et al. [46] | PCA-LDA | The authors have demonstrated all characteristics of the models and loadings shape. | The model is correct. |
| 40 | A. Falamas et al. [47] | PCA-LDA | Ten spectra were recorded for each sample. The authors did not describe in detail the data regarding the chemometrics algorithms and the CV procedure applied, thus it is questionable why ten PCs were selected in the model. | There are no detailed data for the proposed model. |
| 41 | X. Zhu et al. [48] | CARS-PLS | CV is used; the method of splitting the spectra into the training and the test sets is described. | The model is correct. |
| 42 | I.C.V.P.Gogone et al. [49] | PCA-LDA | CV is used; PC loadings are demonstrated. | The model is correct. |
| 43 | E. Ryzhikova et al. [50] | SVM-DA, ANN (MLP) | The model is adequate; however, the description of the utilized methods is presented in the Results section instead of the | The model is correct. |

| | | | Materials and Methods section. | |
|---|---|---|---|---|
| 44 | Z. Gabazana *et al.* [51] | PLS-DA | The model is adequate: CV and VIP analysis are used; however, the VIP scores are not presented. | The model is correct. |
| 45 | N. Chaudhary *et al.* [52] | PCA-LDA | The model is adequate: CV and the optimal number of PCs are used to train the model. | The model is correct. |
| 46 | X. He *et al.* [53] | PCA, BDA, MLP, RBFNN | There is no detailed description of the utilized statistical algorithms; CV is used, the spectra are split into the training and the test sets; however, the authors did not explain how they selected the number of PCs utilized. | There are no detailed data on the proposed model. |
| 47 | M. M. Hassan *et al.* [54] | UVE-PLS | CV is used and RMSECV is provided; however, the PC loadings are not shown. | The model is correct. |
| 48 | F. Chen *et al.* [55] | DOSC, SPA, PLSR, DS | RMSE for prediction set is provided; however, the PC loadings are not shown. | The model is correct. |
| 49 | C. Wichmann *et al.* [56] | PCA-LDA | 50 spectra were recorded for one cell. The CV is randomized, which leads to the spectra of the same cell occurring both in the training and the test sets. There is no explanation of the rule used to select ten PCs in the model. | The model is probably incorrect. |
| 50 | X. Zheng *et al.* [57] | PCA-LDA | The authors did not show the data to prove the selection of the correct number of the PCs utilized. There is no information regarding the cross-validation results. | The model is probably incorrect. |
| 51 | D. Hu *et al.* [58] | PCA-LDA | Three spectra were recorded and analyzed from each sample and the LOPO CV was used. To determine the optimal number of PCs for model training, statistical tests | The model is correct. |
|    |                               |                  | were used, however, the authors did not specify which statistical tests they utilized. There is no indication of the axes on the scores, which makes it difficult to understand which PCs are more informative to classify different samples.                                                                                        |                                                                                                 |
| 52 | N. E. Dina *et al*. [59]      | PCA-Fuzzy LDA    | There is no detailed description of the classification model and the CV method utilized. Decrease in accuracy from 100% to 50% after using the CV is most likely explained by model overfitting                                                                                                                                      | There are no detailed data on the proposed model. The model is probably overfitted.             |
| 53 | C. Marina-Montes *et al*. [60]| PCA              | There is no detailed description of the utilized statistical methods, no validation check, no PCs loadings shape.                                                                                                                                                                                                                    | There are no detailed data on the proposed model.                                               |
| 54 | M. Kopec *et al*. [61]        | PLS-DA           | It is difficult to understand the statistical modeling, because the authors refer to the previous work that studied different experimental samples.                                                                                                                                                                                  | There are no detailed data on the proposed model.                                               |
| 55 | H. F. Nargis *et al*. [62]    | PCA PLS-DA       | There is no explanation of the rule used to select the 18 and 14 PCs in constructed models. There is no indication which CV was used. 15 spectra were recorded for one sample, therefore, the spectra of the same sample may be both in the training and the test sets. The results on the applied CV are not provided. The model can be invalid. | The model is probably incorrect.                                                                |
| 56 | K. Bērziņš *et al*. [63]      | PLSR             | The model validity was checked using an external independent test set. The PC loadings shape and PCs scores value were estimated for building different                                                                                                                                                                              | The model is correct.                                                                           |

| | | | regression models. | |
|---|---|---|---|---|
| 57 | M. A. Bakkar *et al*. [64] | PCA, PLSR | 15 spectra were recorded for each sample that leads to the spectra of the same sample occurring both in the training and the test sets. Ten PCs were used in the model that probably was not correct because, according to the elbow rule, the optimal number of PCs is two or three. Moreover, to determine the optimal number of PCs, it is also necessary to estimate the PCs shapes. The data presented in the manuscript and in Fig.6 in manuscript (Ref. [64]) are different regarding the number of PCs utilized. | The model is probably incorrect. |

**ANN** – Artificial Neural Networks; **BDA** – Bayes Discriminant Analysis; **BPNN** – Back-Propagation Neural Network; **CARS** – Competitive Adaptive Reweighted Sampling; **CNN** – Convolution Neural Network; **CRT** – Classification Regression Tress; **CV** – Cross Validation; **DA** – Discriminant Analysis; **DASN** – Deep architecture-search network; **DS** – Deviation Standardization; **DOSC** – Direct Orthogonal Signal Correction; **HCA** – Hierarchical Cluster Analysis; **GA** – Genetic Algorithm; **kNN** – k Nearest Neighbor; **LDA** – Linear Discriminant Analysis; **LOO CV** – Leave One Out Cross Validation; **LOPO CV** – Leave One Patient Out Cross Validation, **LOSO CV** – Leave-One Sample-Out Cross Validation; **LV** – Latent Variable; **MCARS** – Modified Competitive Adaptive Reweighted Sampling; **MCNN** – Multi-Scale Convolution Neural Network; **MLP Network** – Multilayer Perceptron Network; **NIPALS** – Nonlinear Iterative Partial Least Squares; **PC** – Principal Component; **PCA** – Principal Component Analysis; **PLS** Analysis – Partial Least Squares Analysis/ Progression On The Latent Structures; **PLSR** – Partial Least Squares Regression; **RBFNN** – Radial Basis Function Neural Network; **RF** – Random Forest; **RMSE** – Root-Mean-Square Error; **SPA** – Successive Projections Algorithm; **SVM** – Support Vector Machine; **SWLDA** – Stepwise Linear Discriminant Analysis; **UVE-PLS** – Uninformative Variable Elimination-Partial Least Squares; **VCPA** - Variable Combination Population Analysis; **VIP** – Variables Important On Projection.

One of the main drawbacks that prevents the readers from correct understanding the obtained results is incomplete description of the proposed models in the Materials and Methods section, and, the other way round, overly-detailed explanations of the model parameters in the Results and Discussion section. For instance, H. Wang *et al*. [37] only briefly mentioned data preprocessing in the Materials and Methods while placing the data regarding the utilized chemometrics techniques in the Results without any details on cross validation outcomes (e.g., the proposed model is not provided with RMSE or RMSECV). Another important point is the absence of complete results regarding model construction. Some of the analyzed papers are lacking such results even in the supplementary files, thus the reader is unable to judge about the correctness of the proposed models. Several manuscripts did not report about the parameters of the proposed models and focused only on qualitative performance and results to classify the studied samples [23, 24, 26, 47, 53, 56, 59]. This makes it impossible to objectively analyze the correctness and validity of the statistical models and the success of the achieved results.

A great variety of Raman spectroscopy applications in developing many different multivariate techniques, however, proper statistical analysis of the RS approach to obtain reliable results for sample detection and determination is normally characterized by certain aspects described below.

**1** *Preprocessing of the RS data.* Comparing the RS acquired from different samples usually requires some standardized processing methods to avoid spectral artifacts such as background removing, noise smoothing and binning. However, since detailed description of this step can be found elsewhere [65], we did not focus our attention on the aspects of preprocessing methods.

**2** *Cross-validation (CV).* Overall performance of the classification (and or regression) model is essentially its prediction error, therefore, more reliable results are achieved for the new data that were not used for model training. Frequently, due to limited resources, such as time and other objective reasons, it is impossible to collect a large data set. CV is a well-known tool to verify the classifier performance when a limited dataset is collected. CV requires splitting the experimental spectral data into several parts. Each part is used to predict the performance of the model based on the estimation of the other parts. Therefore, using the CV method makes it possible to check the accuracy and reliability of the results for independent data [66-68]. Therefore, statistical models with a predictive performance determined by the same spectral data that were used to train the statistical model will certainly be overfitted and most likely invalid.

**3** *Splitting into the training and the test sets with regard to CV.* One of the main aspects of a reliable statistical analysis is correct splitting of the spectral data into the training and the validation sets. True diagnostic capabilities of chemometrics modeling need to be estimated using an external independent test set to make predictions by means of new, unseen data at best or by using the spectra of the sample/patients that were not included into the training set.

**4** *Determination of the optimal number of principal components (PCs) and latent variables (LVs).* Determination of the optimal number of PCs is another key factor of building a correct and valid statistical model to analyze the experimental dataset. First of all, the selection of the correct number of PCs can be based on different approaches: the Bartlett chi-square test, the elbow rule (or the scree test) and many others [69]. However, the appropriate number of PCs for the analysis should also be determined by means of certain dataset features [70]. Raman spectral dataset analysis may involve estimating the shape of PCs (or loadings) to make sure that they contain useful information on the chemical composition of the tested samples. In addition, an appropriate number of PCs can be determined on the basis of cross-validation according to the first local minimum in the root-mean square error (RMSE) plot [71].

Now, let us consider some specific features of the statistical models presented in the cited papers to highlight some drawbacks making these models invalid and incorrect. Several papers [13, 2, 5, 32, 47, 56] are missing information regarding CV results. As noted above, CV prevents model overfitting because the training sample is independent from the validation sample. The results obtained without CV are probably over-optimistic and their validity may indeed be insignificant.

The key concern of the majority of the reviewed manuscripts is the number of PCs/LVs in PCA-LDA and PLS-LDA methods utilized. In some of the manuscripts, the description of the rule applied to select the optimal number of PCs/LVs [18, 30, 31, 34, 47, 56, 58, 61] is missing at all. In other manuscripts, the methodology of selecting the correct number of PCs/LVs raises doubts.

For example, the paper by S. Zhu *et al*. [33] demonstrated the results of mice blood serum using SERS measurements to find out the PLS-DA-based biochemical differences between the healthy mice and the mice infected with *C. neoformans*. The authors used the PLS components with a statistically significant difference ($P < 0.05$) that were selected by means of tatistical analysis methods: two sample t-tests. However, using 10 PLS components that were employed to train statistical models and entered into the same LDA model for the analysis of 150 SERS spectra can lead to model overfitting. Assessment of PLS loadings (see Fig. 5 in Ref. [33]) shows that maximal amplitude of PC3, PC4, PC5 loadings is 1/10 of the first and the second loadings, whereas the maximal amplitude of PC6-PC10 loadings is 1/20 and, consequently contains only noise. The 3D scatter plots of the PLS scores also demonstrate that PC1 and PC2 were able to split and distinguish the normal and the infected cases on the plot, whereas PC4 did not demonstrate this ability. As for PC5-PC10, their scores are not provided and there is no way to estimate their ability to distinguish the samples.

In several works [41, 3, 43, 64], the number of PCs is determined by the minimum root mean square error (RMSE) during CV. However, this criterion may be unstable. Despite the fact that increased PC number leads to a minimized error (RMSE) and increased accuracy of the classification model, the RMSE in cross-validation starts varying (in comparison to the RMSE for the obtained model). Fig. 1 demonstrates the results of RMSE estimation for the obtained classification model and the 10-fold cross-validation (the model is constructed using the Raman analysis of human skin to determine kidney failure [11]). To avoid overestimation of the constructed classification model, only the first three PCs should be utilized here.

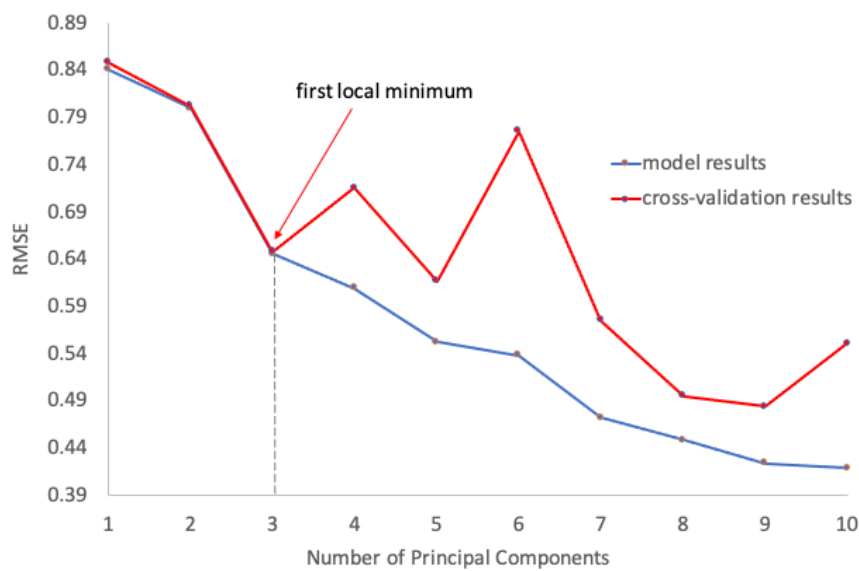

Fig. 1 An example of utilizing RMSE for determining the number of PCs in the PLS model to discriminate a group of patients with kidney failure and a group of healthy volunteers in the analysis of human skin spectra [11]. Reprinted from Comment on "Combining derivative Raman with autofluorescence to improve the diagnosis performance of echinococcosis", Spectrochimica Acta Part A: Molecular and Biomolecular Spectroscopy, Vol 252, I.A. Bratchenko, L.A. Bratchenko, 119541., Copyright (2022), with permission from Elsevier [71].

Papers by Lin *et al*. [12, 28] present an interesting approach to select the PC number. The authors applied the mean squared error of prediction (MSEP) as a more statistically value method for selecting the number of components in PLS by means CV [28]. N components are selected when the following condition is met:

$$\frac{MSEP_N - MSEP_{N+1}}{MSEP_N} \times 100\% < 5\%.$$

Following this approach, three PCs (see Fig. 5(A) in Ref. [28]) were selected for the PLS model that correspond to the non-minimum of MSEP that can lead to a more accuracy value, but the PCs number with a statistically value that will prevent model overfitting.

Although most papers estimate the first local minimum of RMSE, to discard the PCs related to possible noise contribution, it is also necessary to simultaneously estimate, the shape of PC loadings. Thus, in Ref. [63], K. Bērziņš and co-authors demonstrated (more specifically in the Supplementary file) how the loading value is decreasing with a higher number of PCs (see Fig. S7 in Ref. [63]). Using the presented PC score value and the shape of loadings, the authors identified the correct number of PCs for different regression models related to chemical composition. According to the data presented in Fig. S7 in Ref. [63], two PCs were selected for building the regression model based on the calibration set to predict fat content in breast milk [63].

Let us consider the work by M. A. Bakkar and co-authors [64]. The authors demonstrated how Raman spectroscopy coupled with PLSR analysis can verify the Sitagliptin contents in the pharmaceutical samples based on the calibration models prepared under laboratory conditions. In this work, they selected the optimal number of LVs in 10 PCs (see Fig. 5 in Ref. [64]). The authors explained the number of the utilized PCs stating that "*The optimal number of latent*

*variables is usually selected to be where minimum RMSECV is reached. However, the risk of over fitting should also be avoided. For this reason, only 10 latent variables are selected, providing a RMSECV of 0.36. It can be seen that, above 5 there is no significant improvement to the model."* Nevertheless, instead of the minimum RMSECV, it is necessary to estimate the difference in the graph RMSE behavior for the model and for the CV data. According to this approach, only first three PCs should be included into analysis. To prove that a particular PC is appropriate for the model, the shape of that PC should be analyzed as well. Using a lower number of the correct PCs will most likely result in the reduced classification model performance, specifically, in the decreased accuracy or predictive capabilities of the statistical model. Therefore, a high level of qualitative success of this methodology, that was achieved by overestimating its capabilities, does not make it reliable tool for analyzing the chemical composition of specimens.

In some manuscripts [37, 39, 21, 5, 36], an excessive PC number is used for the proposed models, which leads to their overfitting. In [37, 39, 21, 5, 36], the authors claim to have applied CV to identify the optimal number of PCs/LVs. However, the possibility to differentiate the samples by means of a smaller number of PCs/LVs (according to the data presented in the manuscripts) indicates that the PC selection rules were used incorrectly. For example, in [37], H. Wang *et al*. reduced 270 blood high-dimensional spectral data of the two species (human and cattle) to seven PCs (the methodology for selecting the correct number of PCs was not provided). However, in accordance with the results shown for the PCA model (see Fig. 3 in Ref. [37]), the first two or three PCs can be used for accurate differentiating the spectral data of the two species. In [39], S. Bashir *et al*. utilized PCA and PLS-DA to check the diagnostic potential of SERS for distinguishing the tigecyclineresistant E. coli (TREC, 6 cases in the score plot) and the tigecyclinesensitive E. coli (TSEC, 13 cases on the score plot) strains. They reported that *"...The cross-validation provided fourteen (14) optimal LVs in this study which were used to train the PLS-DA model on the calibration dataset..."*. On the other hand, they presented a 2D score plot with LV1 and LV2 indicating a clear difference between the TREC and the TSEC strains. Therefore, the points presented in the paper are contradictory, which questions the validity of the proposed statistical model.

Another important aspect of applying the chemometrics methods to the analysis of spectral data is correct splitting the spectral data into the training and the test sets. A review of the manuscripts presented in Table 1 has shown that in many works the number of spectral measurements significantly exceeds the number of the samples. Due to the limited number of the specific cases, several spectral measurements from a few to hundreds were performed for one sample. Random division of the spectral measurements into the training and the test sets results in an invalid statistical model because this implies that the model will be tested with the spectral data of the same samples that were used to build this model [15, 16, 19, 20, 40, 42, 45, 56, 64]. In this case, with the new spectral data, the proposed model may behave unpredictably. Nargis *et al*. [62] divided their Raman data into the training and the test datasets and obtained the 0.94 ROC AUC for the proposed models. An interesting point that highlights how such performance can be achieved is that the authors analyzed only tens of subjects (29 serum samples were analyzed by Nargis *et al*.). At the same time, in their study, the authors registered 15 spectra from each sample. After recording the spectra, the complete spectral dataset was randomly divided into the training and the test datasets. Therefore, it is quite possible that in the training and the test datasets there will be spectra recorded from the same sample. Taking into account the fact that Raman spectral variability for one tissue or biofluid sample is not statistically significant, most likely, the Raman spectra collected from the same sample are very similar to each other. In this regard, the models proposed by Nargis *et al*. already contain information on the data presented in the test dataset. In these circumstances, it is not surprising that even the overestimated models may demonstrate high performance.

A more suitable approach, when several measurements were recorded from one sample, is provided by LOPO CV/LOSO CV ('leave one patient out/leave one sample out'), when observations of one patient/specimen are excluded, one at a time, from the training set, and the resulting model is evaluated by using the left out observations as tests [17, 25, 58]. A good example of proper splitting the experimental data into the training and the test sets was demonstrated in [12, 63]. Thus, S. Gao and co-authors [12] divided the experimental data (one patient was characterized by the same spectra) into two parts (the training set and the testing set) and reported of high classification results for the PLS-DA model using 10-fold cross-validation and for an unknown testing set. The paper by N. M. Ralbovsky *et al*. [44] shows a statistical algorithm for separating two rat groups (the rats with a standard diet and a high-fat diet that initiated the pre Alzheimer's disease state) that was built and trained using the calibration dataset. However, the general diagnostic capabilities of the algorithm were estimated using external validation to make predictions by the new, unseen data. Therefore, this approach has a better chance that the model is not overestimated and the results are more credible.

Several studies [5, 10, 19, 35, 39, 60] presented the results based on too little experimental data to be considered them as valuable or significant. Good performance obtained by these classification models is explained by the function of the training sample size and can be masked by random testing through a limited sample size [72].

Nevertheless, numerous papers [4, 12, 17, 22, 28, 38, 44, 46, 48-52, 54, 55, 58, 63] have demonstrated adequate results owing to reliable statistical models. Their reviews were added with comments regarding some important aspects of the combined application of Raman technologies and chemometrics, with aspiration that many of these innovative approaches can ultimately reach the cutting edge technologies in practical application, especially for medical diagnostics and clinical practice [73].

**4. Conclusions**

Combination of RS and chemometrics has shown a great potential for creating a universal method for sample chemical analysis and especially for biological and biomedical tasks. This review highlights some important application aspects of chemometrics techniques for Raman analysis datasets that can lead to reliable results. However, we found out that about 70% of the manuscripts analyzed in this review might contain unsupported or invalid data since they their methods are not described and the classification models are deficient.

The fundamental aim of statistical modeling and building classifiers/regressions is not just to obtain the best accuracy and to find the model error; the achieved results must be supported by high-quality, reliable and correct modeling that is primarily based on true chemical features and differences of the studied samples, specimens or structures. If we want the recognition and diagnostic technologies based on the methods of statistic modeling to yield reliable and reproducible results, the following requirements should be observed:

1. a correct and full explanation of the utilized techniques, a detailed description of statistical modeling and all parameters in the Materials and Methods section; if it is impossible to present all the results in the Results section, they should be included into a Supplementary file. It will allow readers to estimate correctness of the proposed classification models;
2. a sufficient experimental sample size for classification/regression to ensure significant and reliable results [72];
3. if the collected data are limited, the classifier/regression performance should be verify by CV [65];
4. correct division of the spectral data into the training and the test/validation sets to prove a true classifier/regression performance of new data;
5. correct selection of the PC number to reduce the dimensionality of the analyzed spectral data and, thus, to avoid model overfitting.

Summarizing, to achieve valid conclusions, research projects and studies should use a sufficient number sample and appropriate statistical analysis techniques [73]. Moreover, the results should be reported accurately and clearly in compliance with the standardized reporting requirements [73], otherwise, it would be impossible to assess the reliability of the proposed approaches. For trustworthy demonstration of practical effectiveness of Raman spectroscopy and chemometrics applied together, we should be focusing on a standardized protocol for techniques and their applications so that task success should not depend on human subjectivity.

## Acknowledgements

The findings of this review were supported by the Russian Science Foundation Project № 21-75-10097.

## Author Contributions

Yulia Khristoforova performed original draft and conceptualization. Lyudmila Bratchenko and Ivan Bratchenko performed editing, conceptualization.